\documentclass{ws-ijmpe}
\begin{document}
\markboth{Brijesh K.Srivastava}{Forward-Backward Multiplicity Correlations in Au+Au Collisions at $\sqrt{s_{NN}}$ = 200 GeV.}
\catchline{}{}{}{}{}

\title{FORWARD-BACKWARD MULTIPLICITY CORRELATIONS IN Au+Au COLLISIONS AT $\sqrt{s_{NN}}$ = 200 GeV.}

\author{{BRIJESH K. SRIVASTAVA} ( for the STAR Collaboration)}

\address{Department of Physics, Purdue University, \\
West Lafayette, Indiana-47907,
USA\\
brijesh@physics.purdue.edu}

\maketitle

\begin{history}
\received{(received date)}
\revised{(revised date)}
\end{history}

\begin{abstract}
The study of correlations among particles produced in different rapidity
regions may provide understanding of the mechanisms of particle production.
 Correlations that extend over a 
longer range are observed in hadron-hadron interactions only at higher energies. Results 
for short and long-range multiplicity correlations (Forward-Backward) are presented 
for Au+Au collisions at $\sqrt{s_{NN}}$ = 200 GeV. The growth of long range correlations are observed as a function of the pseudorapidity gap in central Au+Au collisions. The Dual Parton model and Color Glass Condensate phenomenology have been explored to understand the origin of long range correlations.

\end{abstract}

\section{Introduction}

Multiparticle production can be described in terms of color strings stretched between 
the oncoming partons. These strings then decay into observed secondary hadrons. In this 
scenario, the strings emit particles independently \cite{capela1,capela2,kaidalov}. 
It has been suggested that long-range correlations (LRC) might be enhanced
in hadron-nucleus and nucleus-nucleus interactions, compared to hadron-hadron scattering at the same energy \cite{capela1,capela2}. The presence  of long range correlations implies the 
existence of multiple inelastic collisions and provides a test of the multiple scattering models. The Color Glass Condensate also predicts the large scale rapidity correlations in heavy ion collisions \cite{larry}. 

 Forward-backward multiplicity correlations have been measured in several experiments, mainly in hadron-hadron collisions, to study short
and long range correlations\cite{uhling,alner,ansorge,aexopoulos,derado,na35,e802}. These correlations have also been studied theoretically\cite{capela2,larry,chou,amelin,armesto,braun,giov,shi,urqmd}. 
The correlation strength is defined by the dependence of the average charged particle multiplicity in the backward hemisphere $\langle N_{b}\rangle$, on the
event multiplicity in the forward hemisphere $N_{f}$, $\langle N_{b}\rangle$=a+$b$$N_{f}$, where a is a constant and b measures the strength of the 
correlation\cite{capela1,capela2}:  
\begin{equation}
$b$ = \frac{\langle N_{f}N_{b}\rangle - \langle N_{f}\rangle \langle N_{b}\rangle}{\langle N_{f}^{2}\rangle - \langle N_{f} \rangle ^{2}}= \frac{D_{bf}^{2}}{D_{ff}^{2}} 
\label{b}
\end{equation}  
 $D_{bf}^{2}$ and $D_{ff}^{2}$ are the backward-forward and forward-forward dispersions respectively. The correlation strength given by ``Eq.(1)'' has the contributions both from short and long range sources. The short range correlation comes mainly from sources such as resonance decay, cluster formation and jets. The long range part can be obtained by giving a large gap in rapidity between the forward and backward hemispheres. It has been argued that the short range component falls exponentially in rapidity.
The STAR detector is most suited for forward-backward multiplicity correlations as it is symmetric about mid rapidity.
This is the first measurement of the correlation strength in nucleus-nucleus collisions at the highest RHIC energy.

\section{Analysis}
The data utilized for this analysis is from year 2001 (Run II) $\sqrt{s_{NN}}$ = 200 GeV Au+Au collisions at the Relativistic Heavy Ion Collider (RHIC), as measured by the STAR (Solenoidal Tracker at RHIC) experiment. The main tracking detector at STAR is the Time Projection Chamber (TPC)\cite{starnim}. The TPC is located inside a solenoidal magnet generating a constant, longitudinal magnetic field. For this analysis, data was acquired at the maximum field strength of 0.5 T. All charged particles in the TPC pseudorapidity range -1.0$<\eta<$1.0 and  $p_{T} >0.15$ GeV/c were considered.  The collision events were part of the minimum bias dataset. The minimum bias collision centrality was determined by an offline cut on the TPC charged particle multiplicity within the range -0.5$<\eta<$0.5. The forward-backward intervals were located symmetrically about midrapidity with the distance between bin centers ($\Delta\eta$) ranging from  0.2 to 1.8 with an interval of 0.2. The $\Delta\eta$ = 0.2, 0.4 and 0.6 are termed as inner gap while outer gaps have  $\Delta\eta$=0.8, 1.0, 1.4, 1.6 and 1.8. An analysis of the data from \textit{pp} collisions at 200 GeV was also performed on minimum bias events using the same quality cuts as in the case of Au+Au.

The centralities used in this analysis account for 0-10, 10-20, 20-30, 30-40, 40-50 \% of the total hadronic cross section. An additional offline cut on the longitudinal position of the collision vertex (vz) restricted it to within ±30 cm from z = 0 (center of the TPC). Corrections for detector geometric acceptance and tracking efficiency were carried out using a Monte Carlo event generator and propagating the simulated particles through a GEANT representation of the STAR detector geometry.

However, an autocorrelation exists if the forward-backward multiplicity measurement and centrality determination are made in the same pseudorapidity region. Therefore, the centrality as determined from $|\eta|<0.5$ is used for outer gaps. The centrality determined from the region $0.5<|\eta|<1.0$ is used for inner gaps. 

In order to eliminate (or at least reduce) the effect of impact parameter (centrality) fluctuations on this measurement, each relevant quantity ($N_{f}$, $N_{b}$, $N_{f}^{2}$,  $N_{f}N_{b}$) was obtained on an event-by-event basis as a function of the event multiplicity, $N_{ch}$, and was fitted to obtain the average values of $\left<N_{f}\right>$,
$\left<N_{b}\right>$, $\left<N_{f}\right>^{2}$, and $\left<N_{f}N_{b}\right>$ \cite{ebye1,ebye2}. This method removes the dependence of the correlation strength b, on the size of the centrality bin. 
Tracking efficiency and acceptance corrections were applied to each event. These were then used to calculate the backward-forward and forward-forward dispersions, $D_{bf}^{2}$ and $D_{ff}^{2}$, binned according to the STAR centrality definitions and normalized by the total number of events in each bin.
Systematic effects dominate the error determination. The systematic errors are determined by varying cuts on the z-vertex $|v_{z}|$ ($<$ 30, 20, and 10 cm), number of fit points on the individual tracks in the TPC and the  minimum and maximum value of the fit range for $\left<N_{f}\right>, \left<N_{b}\right>, \left<N_{f}\right>^{2}$, and $\left<N_{f}N_{b}\right>$. The total error is due to both statistical and systematic errors. 

\section{Results and Discussions}

\begin{figure}[th]
\centerline{\psfig{file=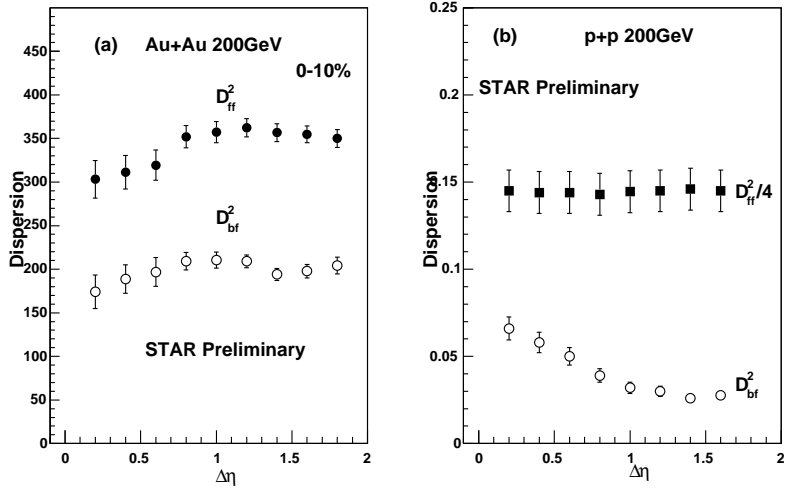,width=11cm}}
\vspace*{8pt}
\caption{ Backward-Forward dispersion ($D_{bf}^{2}$) and  Forward-Forward dispersion ($D_{ff}^{2}$)  as a function of pseudorapidity gap $\Delta\eta$ (a). For Au+Au collisions at $\sqrt{s_{NN}}$ = 200 GeV and (b). For \textit{pp} collisions}
\label{CentralAu}
\end{figure}

\begin{figure}[th]
\centerline{\psfig{file=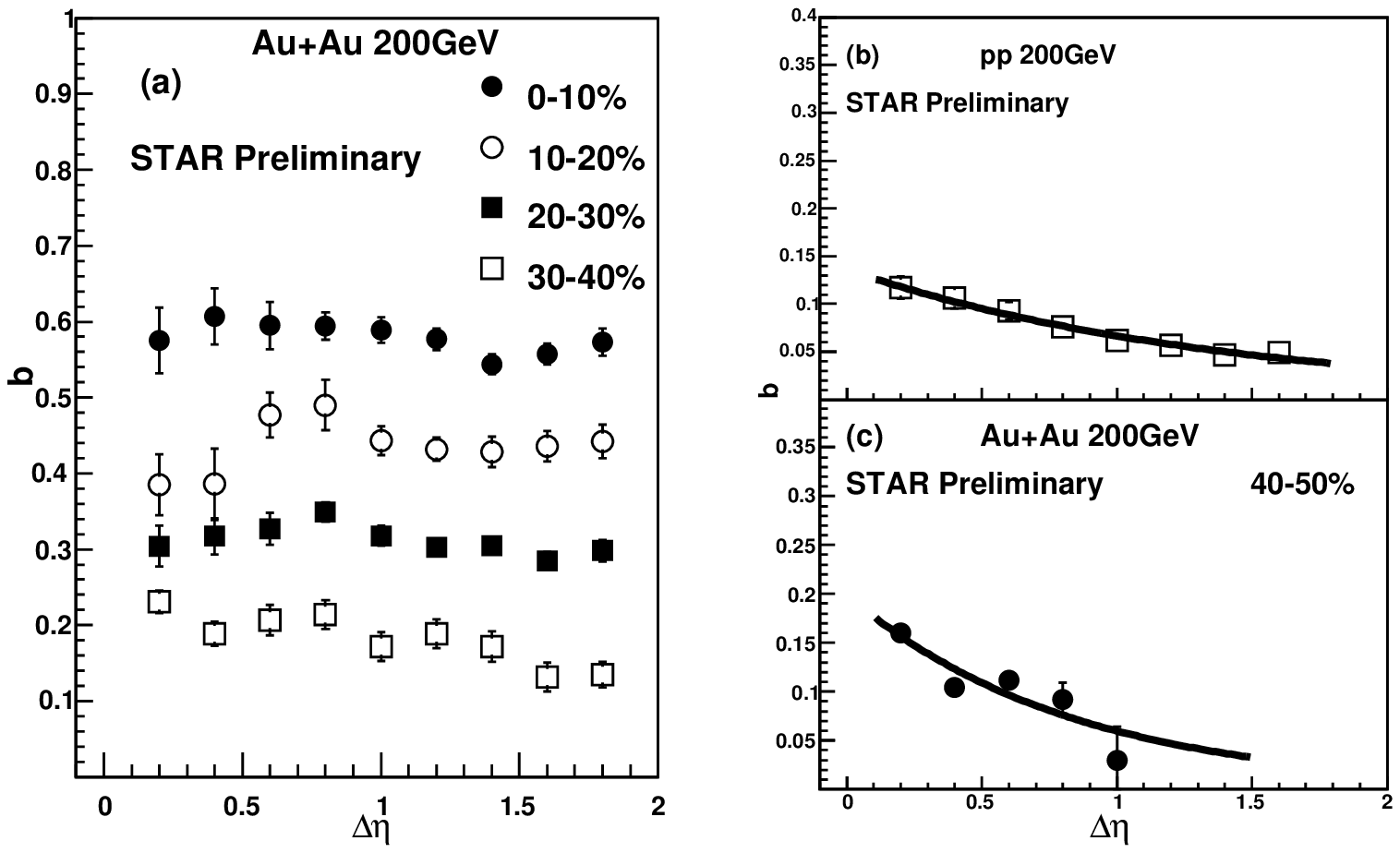,width=15cm}}
\vspace*{4pt}    
\caption{ Correlation strength b as a function of $\Delta\eta$ (a) for Au+Au at four centrality  bins. (b). for \textit{pp}. (c). for 40-50\% Au+Au.}
\label{PeripAuandpp}
\end{figure}

The plot of $D_{bf}^{2}$ and $D_{ff}^{2}$ as a function of the pseudorapidity gap is shown in Fig. \ref{CentralAu}(a) for the 0-10\% most central Au+Au events. It is observed that the value of $D_{bf}^{2}$ and $D_{ff}^{2}$ does not change with the pseudorapidity gap. One can understand the essentially constant value of $D_{ff}^{2}$ with $\Delta\eta$, as it represents the dispersion within the same $\eta$ window, which has approximately the same average multiplicity for all $\Delta\eta$ values.  Fig. \ref{CentralAu}(b) shows $D_{bf}^{2}$, and $D_{ff}^{2}$, as a function of $\Delta\eta$ for the \textit{pp} collisions. Figs. \ref{CentralAu}(a) and (b) show that change in  $D_{bf}^{2}$ with $\Delta\eta$ is quite different in 0-10\% Au+Au as compared to \textit{pp}.  $D_{bf}^{2}$ falls with $\Delta\eta$ for \textit{pp} and can be approximated by Gaussian or exponential function. 
 The $b$ values from  Au+Au and \textit{pp} are shown in  Figs. \ref{PeripAuandpp}(a) and (b) respectively. 
 The $b$ value from \textit{pp} decays exponentially with increasing $\Delta\eta$ as reflected in  $D_{bf}^{2}$(Fig. \ref{CentralAu}(b)). Figs. \ref{PeripAuandpp}(a) and (b) show that the correlation strength with $\Delta\eta$ is quite different in 0-10\% Au+Au as compared to \textit{pp}.

Short range correlations have been extensively studied. The shape of the SRC function has a maximum at, and is symmetric about, mid-rapidity (pseudorapidity, $\eta$ = 0). It can be fitted with an exponential or Gaussian function\cite{capela1,ansorge}:

\begin{equation}
C_{s}(\Delta\eta) \propto exp [-(\Delta\eta)/\lambda] 
\label{SRC1}
\end{equation}
\begin{equation}
C_{s}(\Delta\eta) \propto exp [-(\Delta\eta)^{2}/\delta^{2}]
\label{SRC2}
\end{equation}

where $\lambda$ is the short range correlation length and is related to Gaussian width $\delta$ by $ \lambda = 2\delta/ \sqrt{\pi}$. Thus the SRC is significantly reduced for $\Delta\eta > \lambda.$
According to ``Eq. (\ref{SRC1})'' the $b$ value will have an exponential decay if only the SRC are present. Fig. \ref{CentralAu}(a) shows that in the case of Au+Au the b value is flat and does not show the exponential fall with $\Delta\eta$. There must be another contribution to $b$ which increases with $\Delta\eta$. Since the analysis is limited to $-1<\eta<1$, the short range component cannot be completely eliminated. The \textit{pp} collisions offer a means to estimate the short range component from Au+Au collisions.

An analysis of the mid-central and peripheral events in Au+Au shows that for the 40-50\% centrality bin the correlation strength $b$ is similar to the \textit{pp} case. 
Traditionally, the LRC is measured with a large gap in the forward-backward window and is based on  experimental observation of charged particle correlations in $p \bar{p}$ at $\sqrt {s_{NN}}=$  200, 500, and 900 GeV \cite{ansorge}. If the shape of the correlation function is given by ``Eq. (\ref{SRC1})'' then the strength of the SRC is reduced considerably with $\Delta\eta $ $\sim$ 2.0 units. Thus the remaining portion of the correlation strength is due to the LRC. 

\begin{figure}[th]
\centerline{\psfig{file=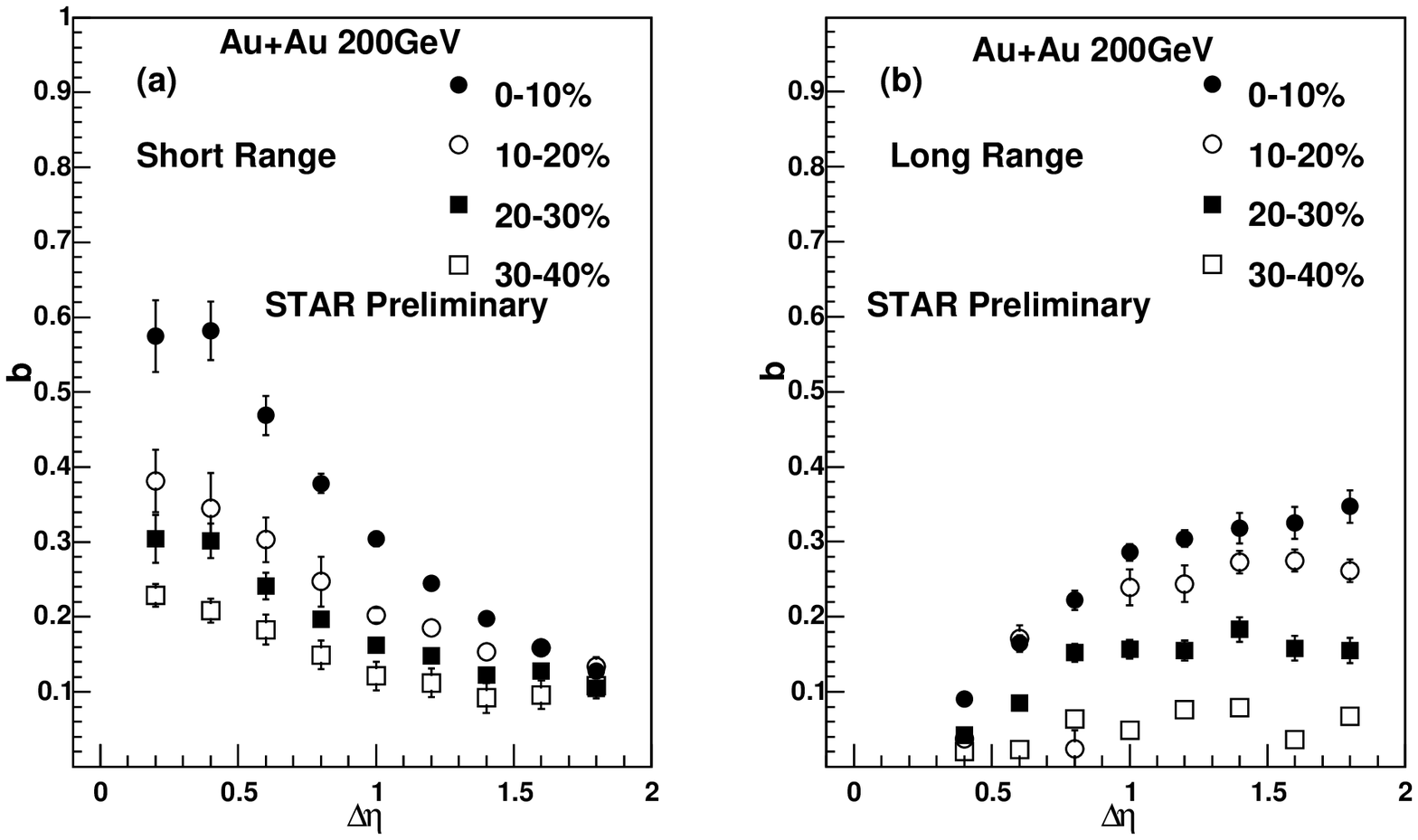,width=11cm}}
\vspace*{8pt}    
\caption{((a). Short range correlation strength as obtained from the scaled b from \textit{pp} as a function of  pseudorapidity gap. (b) Growth of long range correlation for mid central Au+Au events. 
}
\label{Centralb}
\end{figure}

\begin{figure}[th]
\centerline{\psfig{file=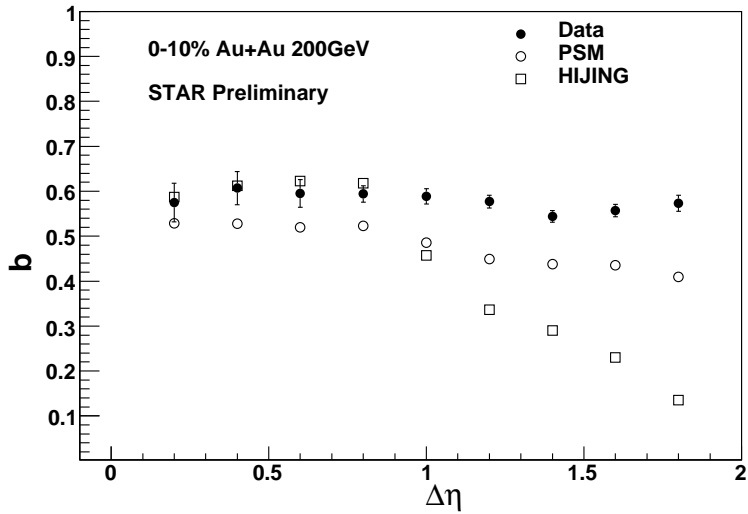,width=11cm}}
\vspace*{8pt}    
\caption{Model comparison with data. The correlation strength is shown for HIJING (open square) and the Parton String Model, PSM(open circle) for the 0-10\% centrality in Au+Au collisions.}
\label{Models}
\end{figure}

An understanding of the interplay between short and long range correlations is described in the framework of the Dual Parton Model (DPM) \cite{capela1}. The numerator of ``Eq. (\ref{b})'' is given by:
\begin{eqnarray}
\nonumber <N_{f}N_{b}>-<N_{f}><N_{b}>  = \\
\nonumber <n>(<N_{0f}N_{0b}>-<N_{0f}><N_{0b}>) \\
+ [(<n^2>-<n>^2)]<N_{0f}><N_{0b}>\hspace{0.05cm}
\label {Dbf}
\end{eqnarray}
where $<N_{0f}>$ and $<N_{0b}>$ are the average multiplicity of charged particles produced in 
the forward and backward hemispheres in each inelastic collision \cite{capela2}. The average number of inelastic collisions is given by $<n>$. 
The first term is the correlation between particles produced in the same inelastic collision, representing the SRC in rapidity. The second term in ``Eq. (4)'', $<n^2>-<n>^2$, is due to the fluctuation in the number of inelastic collisions. This is identified with the LRC \cite{capela1,capela2}.
To get the LRC, the short range contribution must be subtracted 
from central Au+Au collisions. 
As shown in Fig. \ref{PeripAuandpp}(a), it appears that \textit{pp} at 200 GeV only have a SRC.
We assume that the slope of the SRC in $\it pp$ is similar to Au+Au. The 40-50\% Au+Au supports this assumption. HIJING calculations are also in agreement with the data, as shown in Fig. \ref{Models}.
To extract the short range component in 0-10\% Au+Au collisions, the \textit{pp} value of $b$ at $\Delta\eta=0$ (non-overlapping measurement windows), is scaled up to the $b$ value in the Au+Au case, as the SRC has a maximum at $\Delta\eta=0$. The plot of $b$ vs $\Delta\eta$  in Fig. \ref{PeripAuandpp}(b) for \textit{pp} was fitted with an exponential function to obtain parameters of the fit function. The scaled b value for the other $\Delta\eta$ points were calculated using the fit function. The short range portion of the correlation strength for 0-10\% Au+Au is shown in Fig. \ref{Centralb}(a).
The remaining part of the correlation strength (long range), is obtained by subtracting  the short range component from the measured correlation strength as indicated by ``Eq. (\ref{Dbf})''. This is shown in Fig. \ref{Centralb}(b) as the long range. There is a growth of the LRC with increasing $\Delta\eta$. 
The signature of LRC has also been seen in two particle correlation measurements. 
It has been shown that the  charged particle pairs in $\Delta\eta$ ( pseudorapidity) and $\Delta\phi$ (azimuth) for  
near side correlations have enhanced LRC for central Au+Au in $\Delta\eta$ compared to peripheral collisions\cite{tom,ron}.

  The centrality of the collision plays an important role in the growth of long range component of the total correlation strength. Data from 10-20\%, 20-30\%, and 30-40\% most central Au+Au collisions have been analyzed, following the same procedure as for 0-10\% centrality, to determine the evolution of the LRC strength. Fig. \ref{PeripAuandpp}(a) shows the total correlation strength b. The LRC is shown in Fig. \ref{Centralb}(b). The magnitude of the LRC is quite large for the most central collisions when $ \Delta\eta >$ 1.0. From Fig. \ref{Centralb}(b) it is clear that the magnitude of the LRC increases from peripheral to central collisions. The SRC portion of the correlation strength is shown in \ref{Centralb}(a) for mid-central bins. 
      The 0-10\% results are also compared with phenomenological models HIJING \cite{hijing} and the DPM \cite{capela2}. Monte Carlo codes HIJING and the Parton String Model (PSM) \cite{capela2,amelin2,armesto3} were used to generate minimum bias events for Au+Au collisions at 200 GeV. The PSM is based on DPM \cite{capela2}. The analysis was carried out in the same manner as with the real data. The PSM was used without the string fusion option. The variation of b with $\Delta\eta$ is shown in Fig. \ref{Models} along with the experimental  value for 0-10\% central Au+Au collisions. 
HIJING predicts SRC with a large value of $b$ near midrapidity in agreement with the data. A sharp decrease in b is seen beyond the $\Delta\eta \sim$ 1.0. PSM has both short and long range correlations and is in qualitative agreement with the data.

In the DPM, the long range component is given by the second term in ``Eq. (\ref{Dbf})''. In this model particle production occurs via string fragmentation. There are two strings per inelastic collision and the long range component of ``Eq. (\ref{Dbf})'' can be expressed as
\begin{equation}
\langle N_{f}N_{b}\rangle - \langle N_{f}\rangle \langle N_{b}\rangle \propto [\langle n^{2}\rangle - {\langle n \rangle}^{2}] {\langle N_{q-\overline{q}}\rangle}_{f}  {\langle N_{q-\overline{q}}\rangle}_{b}
\label{LRC} 
\end{equation}   
where the average multiplicities of $q-\overline{q}$ in the forward and backward regions is given by $ {\langle N_{q-\overline{q}}\rangle}_{f}$ and  $ {\langle N_{q-\overline{q}}\rangle}_{b}$ respectively in each elementary inelastic collision. ``Eq. (\ref{LRC})'' shows that the LRC is due to fluctuations in the number of elementary inelastic collisions. It is believed that the experimental observation of the LRC originates from these multiple partonic interactions. 

The Color Glass Condensate (CGC) provides a theoretical QCD based description of multiple string interactions \cite{venu,iancu}. The CGC  argues for the existence of a LRC in rapidity, similar to those predicted in DPM. In CGC the long range correlation strength has the form \cite{larry2}:

\begin{equation}
\sigma_{FB}= \frac{1}{1+c\alpha_{s}^{2}}
\end{equation}  
where $\alpha_{s}^{2}$ is to be evaluated at the saturation momentum associated with the centrality of the collision. It clearly becomes more correlated as the centrality increases. This also predicts that the parameter c, which is related to the strength of soft correlated emission, is an increasing function of $\Delta\eta$. It should be noted that forward-backward correlations are also predicted in models incorporating some kind of collectivity e g. percolation. An attempt has been made to correlate the LRC with the percolation density parameter \cite{tjt2}.  

\section{Summary}
In summary, this is the first work on the measurement of the long-range correlation strength ($b$), in ultra relativistic nucleus-nucleus collisions.   
The DPM and CGC argue that the long range correlations are due to multiple parton-parton interactions. This indicates that dense quark-gluon matter is formed in mid-central and central Au+Au collisions at $\sqrt{s_{NN}}$ = 200 GeV. An analysis of the correlation for baryons is in progress to address the CGC/string picture. Further study is being carried out to map the LRC with  the percolation density parameter to obtain the critical value of percolation density at which the LRC is considerably diminished.


\begin{thebibliography}{0}

\bibitem{capela1} A. Capella and A. Krzywicki, 
Phys. Rev. D{\bf 184}, 120(1978).
\bibitem{capela2} A. Capella {\it et al.},
Phys. Rep. {\bf 236}, 225 (1994).
\bibitem{kaidalov} A. B. Kaidalov and K. A. Ter-Martirosyan,
 Phys. Lett. B{\bf 117}, 247(1982).
\bibitem{larry} Y. V. Kovchegov, E. Levin and L. McLerran,
Phys. Rev. C{\bf 63}, 024903(2001).
\bibitem{uhling}
  S. Uhlig {\it et al.},
  Nucl. Phys. B{\bf 132}, 15 (1978).
  
\bibitem{alner}
G. J. Alner {\it et al.}, Phys. Rep. {\bf 154}, 247 (1987).
%
\bibitem{ansorge}
  R. E. Ansorge {\it et al.} ,
  Z.  Phys.  C{\bf 37}, 191(1988).
%
\bibitem{aexopoulos}
T. Alexopoulos {\it et al.}, Phys.  Lett. B{\bf 353}, 155(1995).
\bibitem{derado}
I. Derado {\it et al.}, Z. Phys. C{\bf 40}, 25(1988).
\bibitem{na35}
J. Bachler {\it et al.} [NA35 Collaboration],
Z. Phys. C{\bf 56}, 347(1992).
\bibitem{e802}
Y. Akiba {\it et al.} [E-802 Collaboration],
Phys. Rev. C{\bf 56}, 1544(1997).


\bibitem{chou}
T. T. Chou and C. N. Yang, Phys.  Lett. B{\bf 135}, 175(1984).
\bibitem{amelin} N. S. Amelin {\it et al.},
Phys. Rev. Lett. {\bf 73}, 2813(1994).
\bibitem{armesto} N. Armesto {\it et al.},
Z. Phys. C{\bf 67}, 489 (1995).
\bibitem{braun} M. A. Braun, C. Pajares and  J. Ranft,
Int. J. Mod. Phys. A{\bf 14}, 2689(1999).

\bibitem{giov} 
A. Giovannini and R. Ugoccioni, Phys. Rev. D{\bf 66}, 034001(2002).
\bibitem{shi}
L. Shi and S. Jeon, Phys. Rev. D{\bf 72}, 034904(2005).
\bibitem{urqmd}
S. Haussler, M. Abdel-Aziz, and M. Bleicher,
nucl-th/0608021.

\bibitem{starnim}
  K.~H.~Ackermann {\it et al.}  [STAR Collaboration],
  Nucl.\ Instrum.\ Meth.\ A{\bf 499}, 624(2003).  
\bibitem{ebye1}
  J. Adams {\it et al.}  [STAR Collaboration],
  Phys.\ Rev.\ C{\bf 68}, 044905(2003).
 \bibitem{ebye2} 
  J. Adams {\it et al.}  [STAR Collaboration],
  Phys.\ Rev.\ C{\bf 72}, 044902(2005). 
\bibitem{tom}
J. Adams {\it et al.}  [STAR Collaboration],
Phys.\ Rev.\ C{\bf 73}, 064907(2006).
\bibitem{ron}
J. Adams {\it et al.}  [STAR Collaboration], nucl-ex/0607003.

\bibitem{hijing}
X. N. Wang and M. Gyulassy, Phys.\ Rev.\ D{\bf 44}, 3501(1991);  
\bibitem{amelin2} N. S. Amelin {\it et al.}, Euro. Phys. J. C{\bf 22}, 149(2001).
\bibitem{armesto3} N. Armesto {\it et al.}, Phys. Lett. 527{\bf B}, 92(2002).
\bibitem{venu}
L. McLerran and R. Venugopalan, Phys. Rev. D{\bf 49}, 2233(1994); 
\bibitem{iancu}
E. Iancu, A. Leonidov and L. McLerran, Nucl. Phys. A{\bf 692}, 583(2001).
\bibitem{larry2}
N. Armesto, L. McLerran and C. Pajares, Nucl. Phys. A{\bf 781}, 201(2007).
\bibitem{tjt2}
T. Tarnowsky (2007), these proceedings.
\end{thebibliography}
\end{document}